\documentclass[natbib]{svjour3}            
\smartqed  
\usepackage{graphicx}
\usepackage{aps-bibstyle}  
\newcommand{\la}{\raise0.3ex\hbox{$<$}\kern-0.75em{\lower0.65ex\hbox{$\sim$}}}
\newcommand{\ga}{\raise0.3ex\hbox{$>$}\kern-0.75em{\lower0.65ex\hbox{$\sim$}}}
\newcommand{\muG}{$\mu$G}

\journalname{Space Science Reviews}
\newcommand{\Msun}{\mbox{$M_{\odot}\;$}}
\newcommand{\heatpar}{\alpha_H}
\def\lsim{\;\raise0.3ex\hbox{$<$\kern-0.75em\raise-1.1ex\hbox{$\sim$}}\;}
\def\gsim{\;\raise0.3ex\hbox{$>$\kern-0.75em\raise-1.1ex\hbox{$\sim$}}\;}
\def\alf{Alfv\'en }

\def\cmc{\rm ~cm^{-3}}
\def\kms{\rm ~km~s^{-1}}

\def\cmc{\rm ~cm^{-3}}

\def \kms {\rm ~km~s^{-1}}

\def\lfl{\rm ~ph~cm^{-2}~s^{-1} }

\begin{document}

\title{Magnetic Fields, Relativistic Particles, and Shock Waves
in Cluster Outskirts}

\titlerunning{Cluster Outskirts} 

\author{Marcus Br\"uggen \and Andrei Bykov \and Dongsu Ryu \and Huub R\"ottgering}

\authorrunning{Br\"uggen et al.} 

\institute{Marcus Br\"uggen \at Jacobs University Bremen, Campus Ring
  1, 28759 Bremen, Germany, \email{m.brueggen@jacobs-university.de}\\
\and
Andrei M. Bykov \at Ioffe Institute for Physics and Technology,
194021
St.Petersburg, Russia, \email{byk@astro.ioffe.ru}\\\
\and
Dongsu Ryu \at Department of Astronomy and Space Science,
Chungnam National University, Daejeon 305-764, Korea,
\email{ryu@canopus.cnu.ac.kr}\\
\and
Huub R\"ottgering \at Leiden Observatory, Leiden University, 
P.O. Box 9513, NL-2300 RA, Leiden, The Netherlands,\email{rottgering@strw.leidenuniv.nl}}

\date{Received: date / Accepted: date}

\maketitle

\begin{abstract}
It is only now, with low-frequency radio telescopes, long exposures with high-resolution X-ray satellites and $\gamma$-ray telescopes, that we are beginning to learn about the physics in the periphery of galaxy clusters. In the coming years, Sunyaev-Zeldovich telescopes are going to deliver further great insights into the plasma physics of these special regions in the Universe. The last years have already shown tremendous progress with detections of shocks, estimates of magnetic field strengths and constraints on the particle acceleration efficiency. X-ray observations have revealed shock fronts in cluster outskirts which have allowed inferences about the microphysical structure of shocks fronts in such extreme environments. The best indications for magnetic fields and relativistic particles in cluster outskirts come from observations of so-called radio relics, which are megaparsec-sized regions of radio emission from the edges of galaxy clusters. As these are difficult to detect due to their low surface brightness, only few of these objects are known. But they have provided unprecedented evidence for the acceleration of relativistic particles at shock fronts and the existence of \muG\, strength fields as far out as the virial radius of clusters. In this review we summarise the observational and theoretical state of our knowledge of magnetic fields, relativistic particles and shocks in cluster outskirts.
\keywords{...}

\end{abstract}

\section{Introduction}

Clusters of galaxies are filled with a dilute ($n_e\leq 0.1$
cm$^{-3}$), hot ($T > 10^7$K) gas, the so-called intracluster medium
(ICM). In this review, we are primarily concerned with regions outside of cluster cores, close to the virial radius ($\sim 1$ Mpc). The ICM contains the majority of baryons in galaxy clusters and
has a temperature close to the virial temperature of the gravitational
potential of the cluster.  

Shocks play a fundamental role in the
evolution of the ICM and affect clusters in two
ways: First, shocks thermalize the incoming gas allowing it to
virialise and providing much of the pressure support in
baryons. Secondly, because these shocks are collisionless features
whose interactions in the hot plasma are mediated by electromagnetic
fields, it is possible for a fraction of the thermal distribution of
particles to be accelerated and transformed into nonthermal
populations of cosmic rays (CRs) through diffusive
shock acceleration.  This process results in a part of the kinetic
energy to be converted to non-thermal components. Non-thermal populations of particles can in principle be observed in the hard X-rays through inverse Compton scattering of the Cosmic Microwave background, in the gamma-rays through hadronic interactions of non-thermal protons with the thermal component and in the radio band through synchrotron radiation of the electrons. The latter has been observed in galaxy clusters in the form of diffuse radio sources. These sources are usually subdivided into two classes, denoted as 'radio haloes' and 'radio relics' (see e.g. \cite{2008SSRv..134...93F} for a review).

Radio haloes are diffuse, low surface brightness ($\simeq 10^{-6}$ 
Jy/arcsec$^2$ at 1.4~GHz), steep-spectrum\footnote{S($\nu$)$\propto 
\nu^{- \alpha}$} ($\alpha > 1$) sources, permeating the central 
regions of clusters, produced by synchrotron radiation of relativistic 
electrons with energies of $\simeq 10$~GeV in magnetic fields with 
$B\simeq 0.5-1\;\mu$G. Radio haloes represent the best evidence for 
the presence of large-scale magnetic fields and relativistic particles
throughout the intra-cluster medium (ICM). They are mostly unpolarized and have diffuse
morphologies that are similar to those of the thermal X-ray emission
of the cluster gas. Their origin and the role that shocks play in their formation are still subject to debate, \citep[see \emph{e.g.,} ][for recent review]{pb08,pbr08,brunettiea09}.
Unlike haloes, radio relics are typically located near the periphery
of the cluster. They often exhibit sharp emission edges and many of
them show strong radio polarization. For some relics there is now good evidence that these electrons are produced by shock waves and we will present this evidence in Sec. 4.


If a significant fraction of the pressure of the ICM is provided by
CRs, this will affect gas mass fraction estimates. Since gas mass
fractions in clusters are used to make inferences about dark energy, a
knowledge about the non-thermal content of galaxy clusters is
important for their use in precision cosmology (e.g. Vikhlinin et al. 2010).

There are three types of phenomena that create shocks in the ICM. The
first class of shocks are accretion shocks: At very large off-center
distances, typically several Mpc, cosmological simulations predict
that intergalactic medium (IGM) continues to accrete onto the clusters
through a system of shocks that separate the IGM from the hot, mostly
virialized inner regions. The IGM is much cooler than the ICM, so
these accretion (or infall) shocks should be strong, with Mach numbers
$M \sim 10 - 100$ (e.g., \citealt{bertschinger84, zeldovich72, miniati00}). As such, they were
suggested as sites of effective cosmic ray acceleration, with
consequences for the cluster energy budget and the cosmic gamma-ray
background.  

The second class of shocks are merger shocks: as subhalos fall into the main clusters, they create moderate-strength shocks that
allow the gas to virialise. If an infalling subcluster has a deep
enough gravitational potential to retain at least some of its gas when
it enters the dense, X-ray bright region of the cluster into which it
is falling, a merger shock is produced. In Sec. 2, we will
discuss some prominent observations of this type of shocks.

The third class of shocks are AGN shocks: Powerful AGN are observed to
inflate large bubbles in the ICM, which may generate shocks within the
central few hundred kiloparsecs \citep{bruggen07,simionescu09}. Like
merger shocks, these shocks have moderate Mach numbers, and have been
detected in cluster centres, such as in Virgo and Hydra A. These shocks are
very important because they convert the energy produced by the central
supermassive black hole into thermal energy and, thus, complete the
feedback loop which is regarded as essential for galaxy evolution
\citep{cattaneo09}. As these shocks are primarily found in cluster centres,
they will not be the subject of this review.


The setup of this review is as follows: In Sec. 2, we introduce some phenomena that we will cover in this review and explain some basic terms. In Sec. 3, we describe our current knowledge of cluster shocks from X-ray and $\gamma$-ray
observations. Then we summarise what we can learn from radio
observations in Sec. 4, including estimates of magnetic fields in cluster
outskirts from recent observations of radio relics. In 
Sec. 5, we review recent results concerning structure formation
shocks from cosmological simulations. In Sec. 6, we then discuss the
origin of magnetic fields and relativistic particles in
shocks. Finally, in Sec. 7, we summarise insights gained in the
last years on the physics of shocks in such dilute environments as
found in cluster outskirts and identify open questions.

\section{Fundamentals}

\paragraph{\bf Shock structure:} Shock waves are usually regarded as a sharp transition between a
supersonic (and super-Alfvenic) upstream flow and a subsonic
downstream flow. In a rarefied hot cosmic plasma the Coulomb collisions are not
sufficient to provide the viscous dissipation of the incoming flow,
and collective effects due to the plasma flow instabilities play a
major role. Hence, cosmological shocks in a rarefied highly ionised
plasma are collisionless. Protons, whose thermal velocity is lower than the shock
velocity, are heated dissipatively at the shock layer. 
However, the equilibration of electrons may differ substantially from that of the ions because of their smaller gyroradii. The amount of collisionless heating of electrons depends on the shock velocity and the Mach number \citep[see e.g.][]{Schwartz_ea88,bu99}. Then the Coulomb equilibration eventually occurs on scales that are orders of magnitude longer than 
the collisionless heating scale length \citep[see e.g.][]{fl97,bpp08}.  
Under such circumstances, the shock structure can be more complex
because electrons and ions can remain out of equilibrium for a
while. Then such a shock consists of an ion viscous jump and an
electron-ion thermal relaxation zone. We will revisit the microphysics of cluster shocks in Sec. 6.



\paragraph{\bf Shock acceleration:} Shocks can also change the momentum distribution of a gas by accelerating a
fraction of the particles to very high energies. This has been
observed in astrophysical shocks on a wide range of scales. Examples
are the shock waves that form when the solar wind collides with
planetary magnetospheres, shocks surrounding supernova remnants in the
interstellar medium and shocks in active, extragalactic sources such
as quasars and radio galaxies. The resulting energies reach up to
about 100 keV in the shock in the Earth's magnetosphere to about 100
TeV in supernova remnants. Particles
gain energy whenever they cross the shock front in either
direction. Scattering processes make the velocity distribution
isotropic with respect to the local rest frame on either side of the
shock front. Consequently, there is a certain probability that a
particle that has just gone through the shock and been accelerated
will diffuse back through the shock to where it came from. Every time
the particle crosses the shock it will gain energy. Moreover, the increase in energy is the same no matter
from which direction the particle crosses the (non-relativistic) shock. Particle passing
back and forth through the shock can thus attain very high
energies. This process is called first-order Fermi (Fermi-I)
acceleration or diffusive shock acceleration \citep[DSA;][]{1977DoSSR.234R1306K, 1977ICRC...11..132A, bell78, 1978MNRAS.182..443B, 1978ApJ...221L..29B, 1983RPPh...46..973D, 1987PhR...154....1B, 1991SSRv...58..259J, 2001RPPh...64..429M}. 
The cosmic rays thus generated can penetrate far into the upstream flow, thus creating a shock precursor. The particles decelerate the
flow and preheat the gas. If the cosmic ray pressure inside the shock becomes comparable to the thermal pressure, the cosmic rays will also modify the structure of the shock. Finally, shocks can efficiently generate strong fluctuating magnetic fields in the upstream region due to cosmic-ray driven instabilities \citep{bell04,boe11}. 

\paragraph{\bf Cluster magnetic fields:} The origin of magnetic fields in galaxy clusters remains unclear (see e.g. \cite{dolag08} for a review). It has been suggested
that they are primordial; i.e., a seed field that has formed prior to
recombination is subsequently amplified by compression and
turbulence. Alternatively, it has been proposed that the
magnetic field is of proto-galactic origin \citep{kulsrud97, rkcd08} or that it had been produced by a cluster dynamo \citep{roettiger99}. Others conclude that magnetic fields can be produced
efficiently in shocks by the Weibel instability \citep{medvedev04} or by small-scale plasma instabilities \citep{schekochihin05}. Finally, it has been suggested that the intergalactic medium has been magnetised by bubbles of radio plasma that are ejected by active
galactic nuclei \citep{furlanetto01}. 
Cluster magnetic fields are the subject of a separate chapter in this review and in this review we will focus on magnetic fields in cluster outskirts only.

\paragraph{\bf Turbulence:} Mergers, the motions of galaxies and the action of AGN produce turbulence in the ICM.
Turbulence may have a deep impact on the physics of the ICM
\citep[see e.g.][]{schekochihin05, lazarian06} and on the properties of non-thermal
components in galaxy clusters \citep[see e.g.][]{brunetti11}.
The presence of turbulent gas motions in the ICM is suggested
by measurements of the Faraday Rotation of the polarization
angle of the synchrotron emission from cluster radio galaxies.
These studies show that the magnetic field in the ICM is
tangled on a broad range of spatial scales \citep[see e.g.][]{vogt05} suggesting the presence of super--Alfv\'enic 
motions in the medium.
Also independent attempts from X--ray observations of a number of
nearby clusters, based on pseudo--pressure maps of cluster cores and on the lack of evidence 
for resonant scattering effects in the X-ray spectra, provide
hints of turbulence in the ICM \citep[see e.g.][]{churazov04}.
Important constraints on the fraction of the turbulent and thermal
energy in the cores of clusters are based on the analysis of line broadening
in the emitted X--ray spectra of cool core clusters
\citep{sanders10}. In the outskirts of clusters, simulations predict that the turbulent pressure support increases relative to the thermal pressure support \citep{vazza10,2010arXiv1001.1170P, nagai11}.





\section{Cluster shocks in X- and $\gamma$-ray emission}

\subsection{Cluster shocks in X-rays}

Regions of high-entropy gas in clusters have been observed by \emph{ROSAT,
ASCA, XMM-Newton and Chandra} and have been interpreted as the result
of shock heating (e.g., \citealt{belsole05}). However, only a handful
of shock fronts, exhibiting both a sharp gas density edge and an
unambiguous temperature jump, have been found: in the "bullet
cluster", 1E 0657Ð56 \citep{markevitch02}, Abell 520 \citep{markevitch05}, 
and Abell 2146 \citep{russell10}. Such discoveries are so rare because
one has to observe a merger when the shock has not yet moved to the
low-brightness outskirts, and is propagating nearly in the plane of
the sky, to give us a clear view of the shock discontinuity. In
addition, the shocks in A520 and 1E 0657Ð56 have Mach numbers of $M\sim 2 - 3$, which
provide a big enough gas density jump to enable accurate
deprojection. Merger shock fronts have been found in other clusters,
e.g., in A3667, \citep{finoguenov10}, A754, \citep{krivonos03}, A521,
\citep{giacintucci08}.
 
Recently, \cite{finoguenov10} analysed \emph{XMM-Newton} and \emph{Suzaku} pointings of the
cluster Abell 3667, including the NW relic (see Fig. 1). They find a sharp X-ray surface
brightness discontinuity at the outer edge of the radio relic, and a
significant drop in the hardness of the X-ray emission at the same
location. This discontinuity is consistent with a Mach number 2 shock
moving at 1200 km/s associated with the merger. Kinetic energy is
being dissipated in this shock at a rate of $dE_{\rm KE}/dt \sim 1.8\times 10^{45}$
erg/s.
The total non-thermal luminosity of the relic is $L_{\rm NT} \approx 3.8 \times 10^{42} [ ( 3.6 \,\mu{\rm G} / B )^2 + 1 ]$ erg s$^{-1}$.  If this energy is provided by the
shock acceleration of electrons, then a fraction $\epsilon\sim 0.2$ \% of the kinetic energy flux through the shock is converted to the non-thermal luminosity.
Shock acceleration at the outer edge of the relic and radiative
losses as the electrons are advected away from the shock can explain
the rapid steepening of the radio spectral index with distance from
this edge of the relic. Alternatively, the surface brightness
discontinuity and hardening of the X-ray spectrum might be due to
Inverse Compton (IC) emission from the relic. In this case, the
magnetic field in the relic is about 3 \muG. Since the observed X-ray
excess from the relic is an upper limit to IC emission, this yields a
lower limit on the relic magnetic field of $\geq$ 3 \muG. This is a
remarkably strong magnetic field at this large projected distance (2.2
Mpc) from the cluster center, but is consistent with Faraday rotation
through the relic observed towards two background radio galaxies.

\begin{figure*}
\includegraphics[width=1\textwidth]{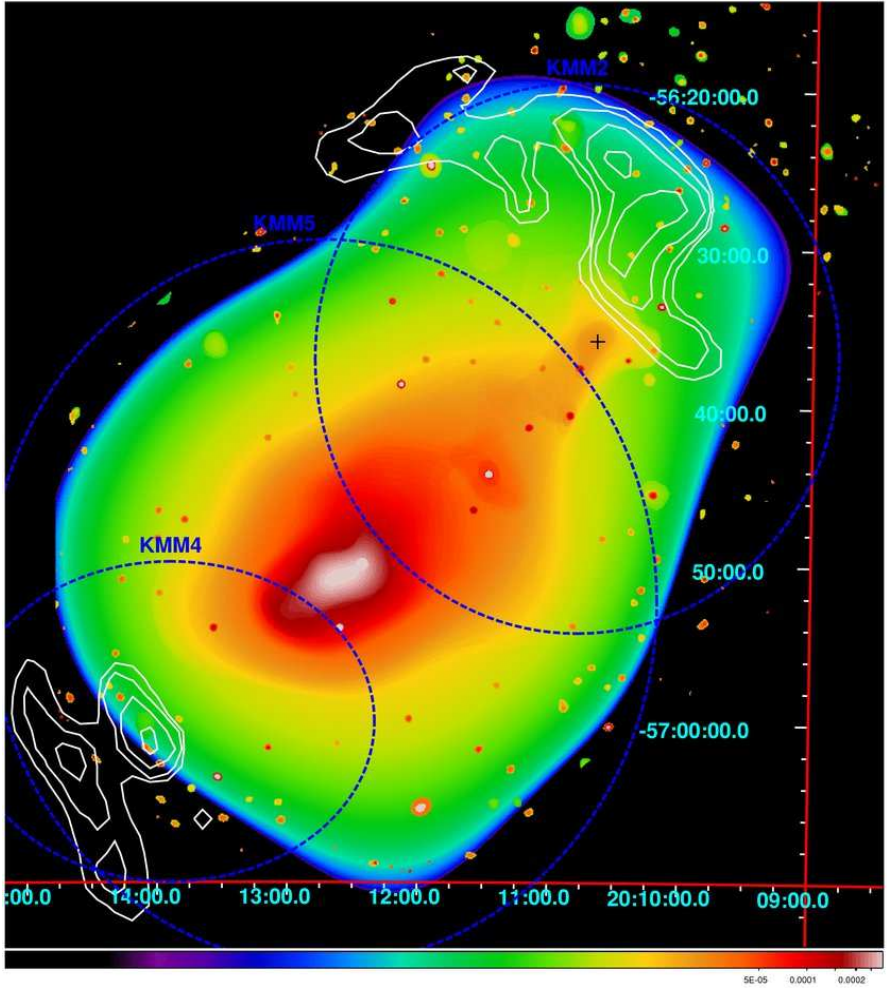}
\caption{Wavelet reconstruction of the 0.5-2 keV X-ray mosaic of A3667
  from \emph{XMM-Newton}. There is a sharp discontinuity in the X-ray surface
  brightness to the NW; the blue ellipses show the location, name and
  size of three main galaxy components (KMM5, KMM2, and KMM4) of A3667
  from the analysis of Owers et al. (2009). The black cross marks the
  center of KMM2 group. The white contours show the wavelet
  reconstruction of the SUMMS 843 MHz radio image in the region of
  radio relics. The contours are drawn at the 1, 3, 9, and 20 mJy/
  beam levels. From Finoguenov et al. (2010) with kind permission.}
\label{fig:3667}
\end{figure*}

The bow shock in 1E 0657Ð56 offers a unique experimental setup to
determine how long it takes for post-shock electrons to come into
thermal equilibrium with protons in the ICM (Markevitch et
al. 2006). Currently, X-ray observations only yield the electron
temperature but not the ion temperature. However, one can use the 
measured gas density jump at the shock front and the pre-shock
electron temperature to predict the post-shock adiabatic and
instant-equilibration electron temperatures, using the adiabatic and
the Rankine-Hugoniot jump conditions, respectively, and compare them
with the data. Furthermore, if the downstream velocity of the
shocked gas flowing away from the shock is known, one can deduce how the flow
spreads out the electron temperature along the spatial coordinate in
the plane of the sky. The Mach number of the 1E 0657Ð56 shock is
sufficiently high, such that the adiabatic and shock electron
temperatures are sufficiently different (for $M \leq 2$, they become
close and difficult to distinguish, given the temperature
uncertainties). Moreover, the distance traveled by the post-shock
gas during the time, is well-resolved by \emph{Chandra}. Markevitch et
al. (2006) found that the temperatures are consistent with instant
heating; equilibration on the collisional timescale is excluded,
although with a relatively low 95\% confidence. The equilibration
timescale should be at least 5 times shorter than the Coulomb time.
We will revisit the theory of dissipation in collisionless shocks in
Sec. 6.

The low surface brightness of the cluster outskirts is challenging for current
X-ray telescopes \citep[e.g.][]{bw10, molendi10}.
Projected temperature profiles to $\sim 0.5
r_{200}$ of 15 nearby $(z<0.2)$ clusters were derived from
\emph{XMM-Newton} observations by \cite{prattea07} \citep[see also][]{vikhlininea05}.

The low and well-constrained background levels make \emph{Suzaku} the
prime instrument for the study of X-ray emission from cluster
outskirts. \emph{Suzaku} is in a low orbit within Earth's magnetopause
providing significantly lower and stable particle background compared
to \emph{Chandra} and \emph{XMM-Newton}. However, the calibration of the instruments can sometimes be difficult. The low surface brightness
outskirts of A2204 around $r_{200}$ were studied by
\citet{reiprichea09} with \emph{Suzaku}. 
\cite{simionescu11} determined ICM properties out to the edge of the Perseus Cluster. Contrary to earlier findings, the cluster baryon fraction is consistent with the expected universal value at half of the virial radius. The apparent baryon fraction exceeds the cosmic mean at larger radii, suggesting a clumpy distribution of the gas for radii larger than half of the virial radius. Also with \emph{Suzaku}, \citet{georgeea09}
studied X-ray emission from the outskirts
of the cluster of galaxies PKS 0745-191 at $z=0.1028$. They determined radial profiles of density, temperature,
entropy, gas fraction, and mass beyond the virial radius out to $\sim
1.5 r_{200}$. The temperature is
found to decrease by roughly 70~\% from
0.3-1~$r_{200}$.  \emph{Suzaku} observations
provide evidence for departure from hydrostatic equilibrium around
and even before the virial radius as it is seen in the galaxy cluster
Abell 1795 \citep[e.g.][]{bautzea09}. \emph{Suzaku}
observations of PKS 0745-191 revealed one shock front possibly indicating that this cluster is still accreting
material, likely along a filament. Evidence for nonthermal pressure
support suggests that bulk motions from merger activity could be
making a significant contribution to the gas energy in the outskirts
of this cluster \citep{georgeea09}. There is some evidence for the
presence of an excess above the dominating thermal emission of the ICM
hot gas in both X-rays and EUV \citep{rephaeliea08,durretea08,ma09}. However, a
recent search by \citet{ajelloea09} of hard X-ray excess above the
thermal emission in a large sample of clusters with \emph{Swift Burst
  Alert Telescope} does not show significant nonthermal hard X-ray
emission.  The only exception is Perseus whose
high-energy emission \citet{ajelloea09} attributed to the central
galaxy NGC 1275. In the next years, as more cluster outskirts are being observed 
with \emph{Suzaku}, more conclusive results on shocks and nonthermal components are expected.

Compared to X-ray observations, the outer regions of clusters can potentially be better studied by Sunyaev-Zel'dovich (SZ) observations because the SZ effect depends only on the electron density to the first power, while the X-ray emission depends on the electron density squared. Facilities such as the Atacama Cosmology Telescope (ACT), the South Pole Telescope (SPT) and the Planck satellite are searching for the SZ signal of galaxy clusters and have found some interesting first results. For example, it was found that the SZ signal caused by clusters is by a factor of $\sim 2$ smaller than predicted by models of clusters that suggest that the pressure by the electrons has been overpredicted \citet{lueker10}. This is presumably caused by a substantial nonthermal pressure at cluster outskirts, most of it is likely to be turbulent pressure \citep[see e.g.][]{shaw10}. This picture is also supported by simulations \citep{vazza10,2010arXiv1001.1170P, nagai11} which we review in Sec. 5.


Given the advancements in the X-ray and SZ observations of the cluster outer regions, as well as the growing evidence of missing thermal energy in the ICM and the possible negative implications for cosmological tests, a more detailed study of the kinetic processes in cluster envelopes is necessary.

\subsection{$\gamma$-ray Observations}

Recently, \citet{ackermannea10} reported on the search for GeV
emission from clusters of galaxies using the \emph{Large Area
  Telescope} on the \emph{Fermi Gamma-ray Space Telescope}. Only upper limits
on the photon flux were reported in the range 0.2-100~GeV toward a
sample of around 30 observed clusters (typical values $(1-5) \times
10^{-9} \lfl$), considering both point-like and spatially resolved
models for the high-energy emission. The authors concluded that the
volume-averaged relativistic hadron-to-thermal energy density ratio is
below $5\%-10\%$ in several clusters. Also, using \emph{High Energy
  Stereoscopic System (H.E.S.S.)} observations of Coma cluster,
\citet{aharonianea09} placed an upper limit $\sim 10^{-13} \lfl$ at
photon energies above 5~TeV for the Coma core of 0.2 degrees in radius
constraining the multi-TeV particle population in the cluster.
In summary, at this time the $\gamma$-ray observations remain
inconclusive with respect to the origin of diffuse radio emission in
clusters, but this is bound to change soon as the upper limits
are expected to decrease.

\section{Cluster shocks in radio emission}

A number of diffuse, steep-spectrum radio sources
without optical identification have been observed in galaxy
clusters. The emission from these sources is synchrotron radiation, which indicates the presence of highly
relativistic electrons and $\mu$G range magnetic fields.  Especially radio relics are thought to trace cosmological shocks. 
Radio relics can be divided into two main groups \citep{2004rcfg.proc..335K}. \emph{Radio gischt} are large elongated, often Mpc-sized, radio sources located in the periphery of merging clusters. They probably trace shock fronts in which particles are accelerated via the diffusive shock acceleration
mechanism. Among them are double-relics with the two relics located on both sides of a cluster center 
\citep[e.g., ][]{bonafede09, 2009A&A...506.1083V, 2007A&A...463..937V, 2006Sci...314..791B, 1997MNRAS.290..577R}. According to DSA, the integrated radio spectrum should follow a single power-law. \emph{Radio phoenices} are related to radio-loud AGN. Fossil radio plasma from a previous episode of AGN activity is thought to be compressed by a merger shock wave which boosts, both, the magnetic field inside the plasma as well as the momenta of the relativistic particles. As a result the radio plasma brightens in synchrotron emission. In contrast to the radio gischt, the phoenices have a steep curved spectrum\footnote{$F_{\nu} \propto \nu^{\alpha}$, with $\alpha$ the spectral index} indicating an aged population of electrons.

The sizes of relics and the distance to the cluster centre vary significantly. Examples for radio
relics with sizes of 1~Mpc or even larger have been observed in Coma
(the prototype relic source 1253 + 275, \citep{giovannini91},
Abell 2255 \citep{feretti97} and Abell 2256 \citep{rottgering94},
which contain both a relic and a halo (as do Abell 225, Abell 521, Abell 754, Abell 1300, Abell 2255 and Abell 2744). The cluster Abell 3667 contains two very luminous, almost symmetric
relics with a separation of more than 5~Mpc \citep{rottgering97}, as does ZwCl 2341.1+0000 \citep{vanweeren09, bonafede10}, and Abell 2345 and Abell 1240 \citep{bonafede09}.
The clusters A115 and A1664 show relics only at one side of the elongated X-ray
distribution \citep{govoni01}. Recently, another double radio relic
was found in the galaxy cluster ZwCl 0008.8+5215
(van Weeren et al. 2010) which we will discuss in detail below.

\begin{figure*}
\includegraphics[width=1\textwidth, angle=90]{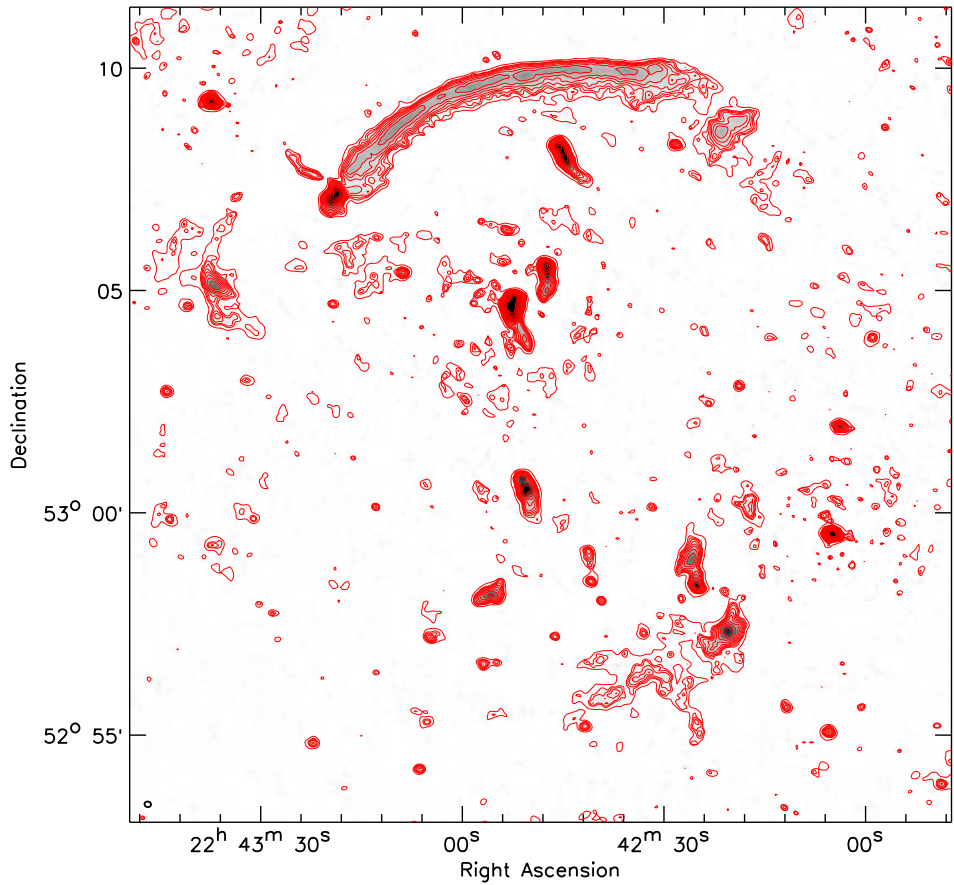}
\caption{Radio map (GMRT 325 MHz) of the relic in CIZA J2242.8+5301. From \cite{vanweeren10}.}
\label{fig:ciza1}
\end{figure*}
\begin{figure*}

      \includegraphics[angle = 90, trim =0cm 0cm 0cm 0cm,width=0.7\textwidth]{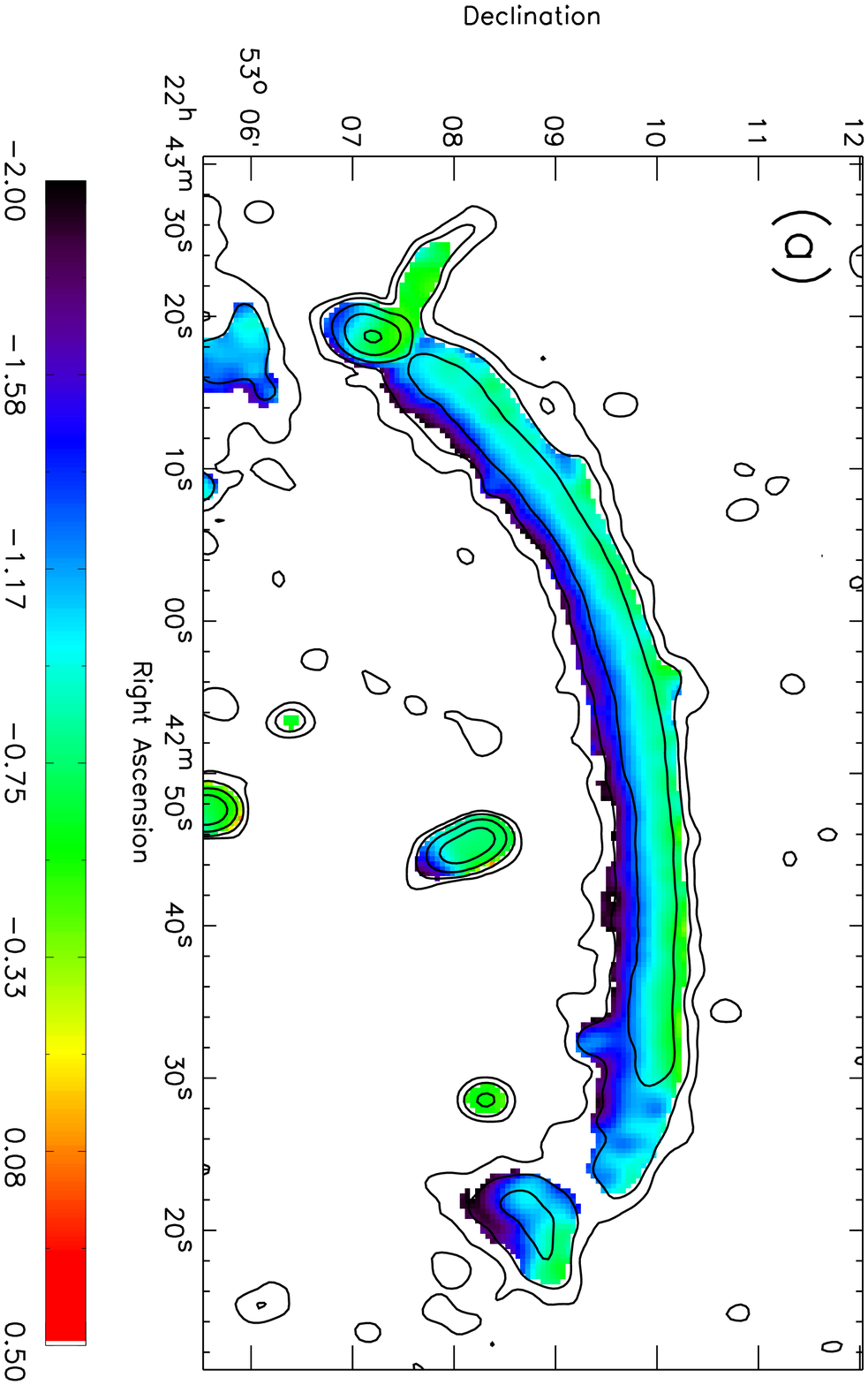}
      \includegraphics[angle = 90, trim =0cm 0cm 0cm 0cm,width=0.7\textwidth]{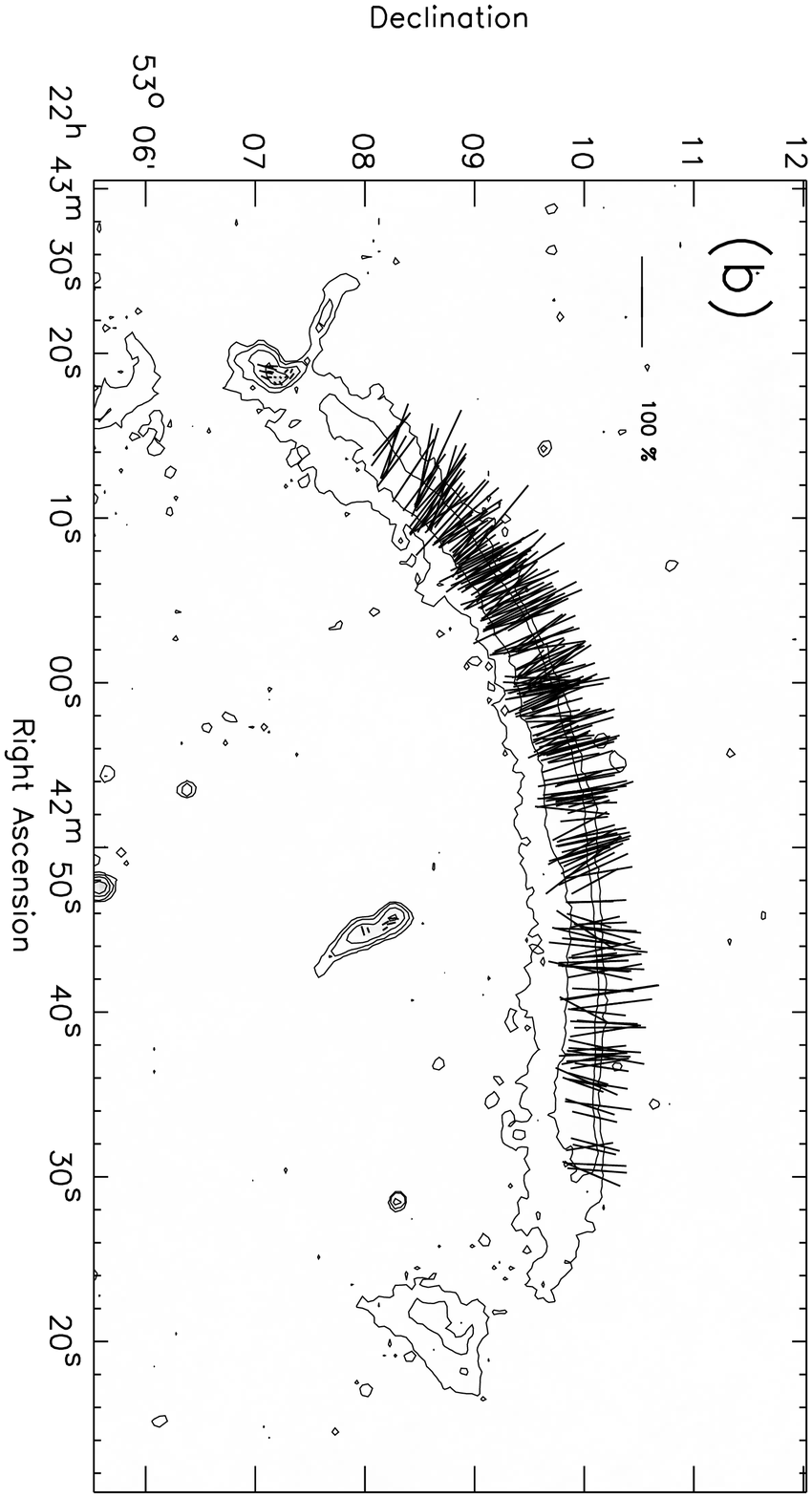}
\caption{Radio spectral index and polarization maps of the relic in CIZA J2242.8+5301. Top: The spectral index was determined using 
matched observations at 2.3, 1.7, 1.4, 1.2, and 0.61 GHz, fitting a power-law radio spectrum to 
the flux density measurements. Bottom: The 
polarization electric field vector map was obtained with the VLA at a frequency of 4.9 GHz. The length of the vectors is proportional the polarization fraction, which is the ratio between the total intensity 
and total polarized intensity. A reference vector for 100\% polarisation is drawn in the top left 
corner. From \cite{vanweeren10}.}
\label{fig:ciza2}
\end{figure*}

The relic with the best evidence for shock acceleration found to date is located in the northern
outskirts of the merging galaxy cluster CIZA~J2242.8+5301
($z=0.1921$), see Fig.~\ref{fig:ciza1}. The relic is located at a
distance of 1.5~Mpc from the cluster center and spans 2~Mpc in length.
The relic shows a clear unambiguous spectral index gradient towards
the cluster center. The spectral index, measured between 2.3 and
0.61~GHz, steepens from $-0.6$ to $-2.0$ across the width of the
narrow relic. The gradient is visible over the entire 2~Mpc length of
the relic, something that has never been observed before. This is a
crucial observation as it constitutes clear evidence for diffusive shock
acceleration and spectral ageing of relativistic electrons in an
outward moving shock. The relic is strongly polarized at the 50-60\%
level, indicating an ordered magnetic field, and polarization magnetic
field vectors are aligned with the relic (see
Fig.~\ref{fig:ciza2}). In the southern part of the cluster a second
fainter broader relic is found and the elongated radio relics are
orientated perpendicular to the major axis of the cluster's elongated
ICM. This is exactly as expected for a binary cluster merger event in
which this second southern relic traces the shock wave that travels in the opposite direction from the first one. Furthermore, a faint halo of
diffuse radio emission is seen extending all the way towards the
cluster center connecting the two radio relics. This emission extends
over 3.1~Mpc, making it by far the largest known diffuse radio source
in a cluster to date.

The spectral index at the front of the relic is $\alpha=-0.6 \pm 0.05$. In simple shock acceleration theory, $\alpha$ is related to the Mach number via $\alpha=-(3+M^2)/(2M^2-1)$ \citep{rosswog07}, which yields a Mach number of $4.6_{-0.9}^{+1.3}$. Subsequent hydrodynamic simulations of this shock front have indicated slightly lower Mach number of around 3. Using the $L_{X}-T$ scaling relation for
clusters \citep{markevitch98}, we estimate the average
temperature of the ICM to be $\sim9$~keV. Behind the shock front the
temperature is likely to be higher. Using the redshift, downstream velocity,
spectral index, and characteristic synchrotron timescale, the width of
the relic is given by

\begin{equation}
l_{\rm{relic}} \approx 110 \mbox{ kpc} \times \frac{B^{1/2} B_{\rm{CMB}}^{3/2}}{B^2 + B_{\rm{CMB}}^{2}}
\end{equation}
for a relic seen edge-on without any projection effects. $B_{\rm{CMB}}$  is the equivalent field strength of the IC scattering from the Cosmic Microwave Background, which is known, and thus the measurement of $l_{\rm{relic}}$ from the radio maps, directly constrains the magnetic field.  From the 610 MHz
image (the image with the best signal to noise ratio and highest
angular resolution), the relic has a deconvolved width (FWHM) of
55~kpc. Taking into account projection
effects, van Weeren et al. (2010) conclude that the magnetic field strength at the
location of the bright radio relic lies between 5 and 7~$\mu$G.
As in A3667, this measurement points towards \muG\,- scale fields in merger
shocks found at large distances from the cluster centre.

\begin{figure*}
\includegraphics[width=1\textwidth]{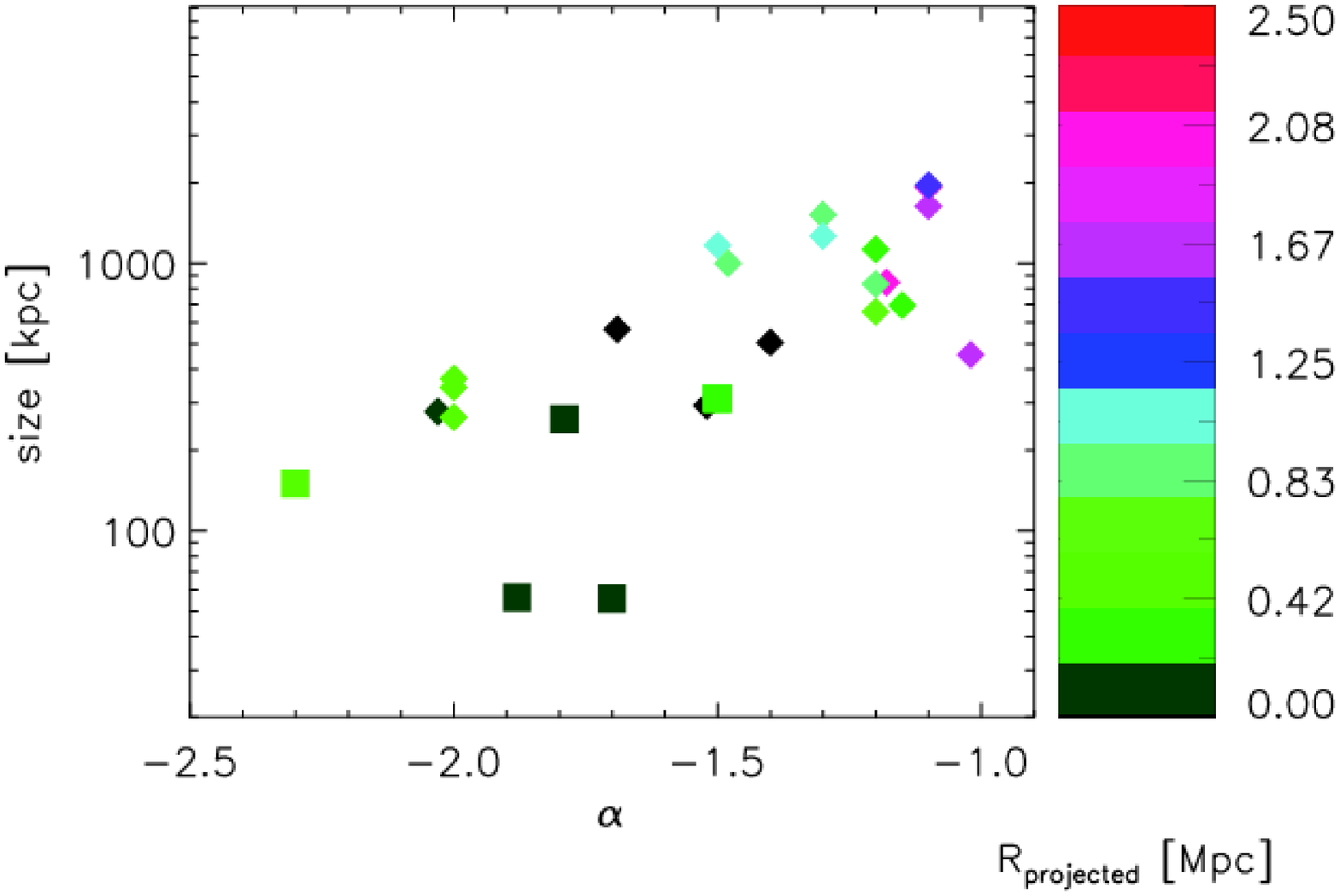}
\caption{Spectral index of radio relics versus their size. Squares are
  the proposed radio phoenices. Diamonds represent
  the radio relics tracing merger shocks where particles are being
  accelerated by shock acceleration. The color coding is according to
  the projected distance from the cluster center. For the relics
  represented by black symbols we could not obtain a reliable
  projected distance to the cluster center. From \cite{2009A&A...508...75V}.}
\label{fig:sizealpha}
\end{figure*}

Even though the sample of known radio relics is still small, one can
start to find correlations between size, location and spectral index
of these unique sources, which can eventually be compared to
simulations of relic formation.
The spectral index of the radio relics, versus the physical size is
shown in Fig.~\ref{fig:sizealpha}. The projected distance from the
cluster center is color coded. Van Weeren et al. (2010) find that on average the smaller
relics have steeper spectra. Such a trend is in line with predictions from
shock statistics derived from cosmological simulations
\citep{skillman08, battaglia09,hoeft08}. They find that larger shock
waves occur mainly in lower-density regions and have larger Mach
numbers, and consequently shallower spectra. Conversely, smaller shock
waves are more likely to be found in cluster centers and have lower
Mach numbers and steeper spectra. We note that more spectral index
measurements of high quality are needed to confirm the correlation
between physical size and spectral index.

\begin{figure*}
\includegraphics[width=1\textwidth]{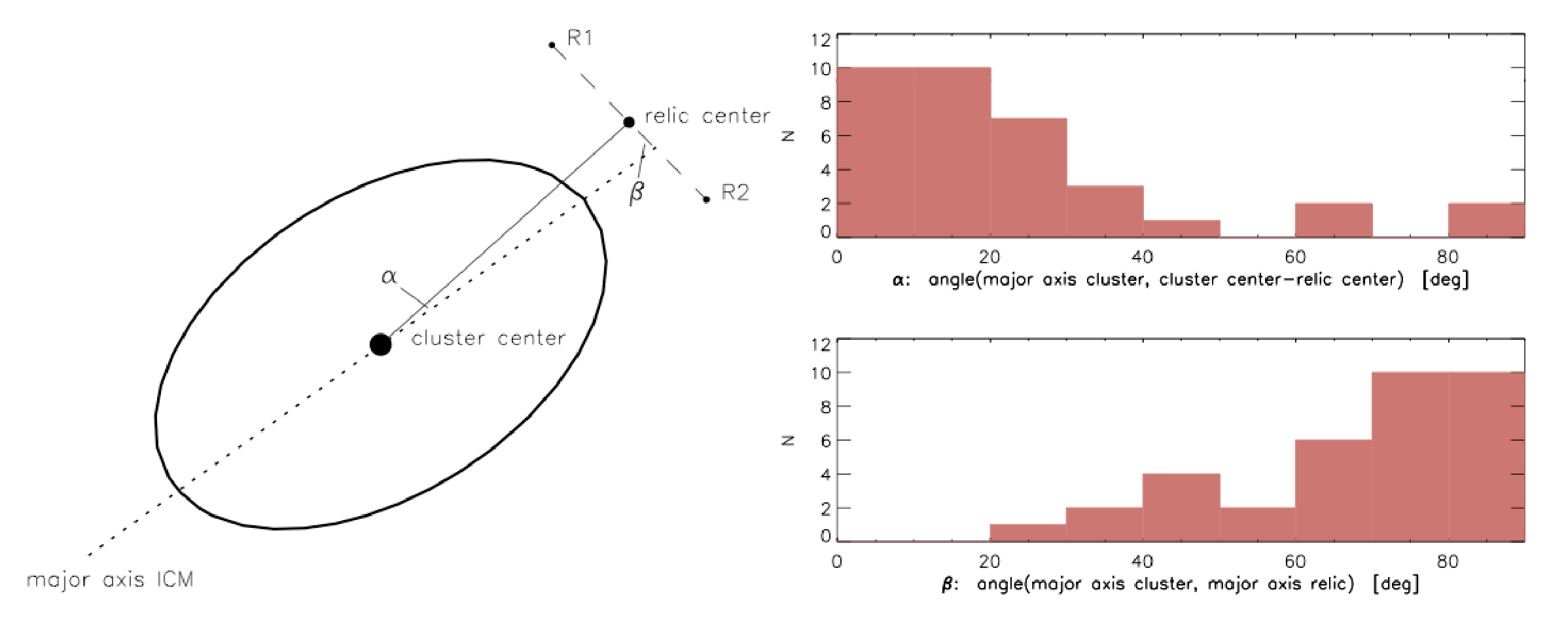}
\caption{Left: Schematic illustration of the angle between the major axis of the ICM and the line relic centerÐcluster center ($\alpha$), and the angle 
between the relic orientation and major axis of the ICM ($\beta$). Top Right: Histogram of angles between the major axis of the X-ray emission and the 
line connecting the cluster center with the center of the relic. Bottom Right: Histogram of angles between the major axis of the X-ray emission 
and the relicÕs major axis. From \cite{vanweeren11}.}
\label{fig:orientation}
\end{figure*}
A further confirmation of the binary merger origin of most radio relics comes from measuring the angle of orientation of the relic's main axis with respect to the major axis of the X-ray surface brightness map of the host cluster. We have computed the angle $\alpha$ between the major axis of the ICM and the line cluster centerÐradio relic center. The resulting histogram is shown in 
Fig. 5. From this histogram we see that relics are preferably 
found along the major axis of the ICM. This is in line with the 
simple picture that shock waves propagate outwards along the 
merger axis. We also calculate the angle $\beta$ between the major 
axis of the ICM and the relicÕs major axis. Fig. 18 shows that most 
relics are oriented perpendicular to the ICM major axis, also in 
agreement with a shock origin for radio relics.

In the next years, simulations should be able to reproduce these
trends, and thus constrain the efficiencies of shock acceleration and
the magnetic field evolution in clusters. At the same time, it is
predicted that \emph{LOFAR} will find of around 100 new radio relics \citep{hoeft08}. 
This will help address some remaining puzzles that surround radio relics, most importantly:

\begin{itemize}
\item Which processes accelerate electrons so efficiently at relatively low Mach number ($M\sim 2-4$) shocks?
\item What produces the magnetic fields inside relics? Both the inferred field strengths as well as the observed polarisation of the radio emission need to be explained.
\item Why do some relics have very sharp edges while others appear very fuzzy?
\item Under which conditions do relics form? When do we see single and when double relics?
\item Some relics appear to be connected to cluster-wide radio halo emission. Is there a physical connection between the two?
\end{itemize}
In the next section, we will review the current state of cosmological
simulations that include a treatment for shock acceleration of cosmic
rays.

\section{What we know from simulations}

\subsection{Shocks in cosmological simulations}

The statistics of cosmological shocks in the large-scale structure of
 the Universe has been derived from simulations using, both, Eulerian
 hydrodynamic codes (e.g. \citealt{miniati00,miniati01,ryu03,krco07,skillman08,vazza10,vazza09a,vazza09b,molnar09})
 and smoothed particle hydrodynamic codes \citep{pfrommer07,hoeft08} as well as a number of
semi-analytical \citep{gabici03,keshet03} studies.
 

\begin{figure*}
\includegraphics[width=1.1\textwidth]{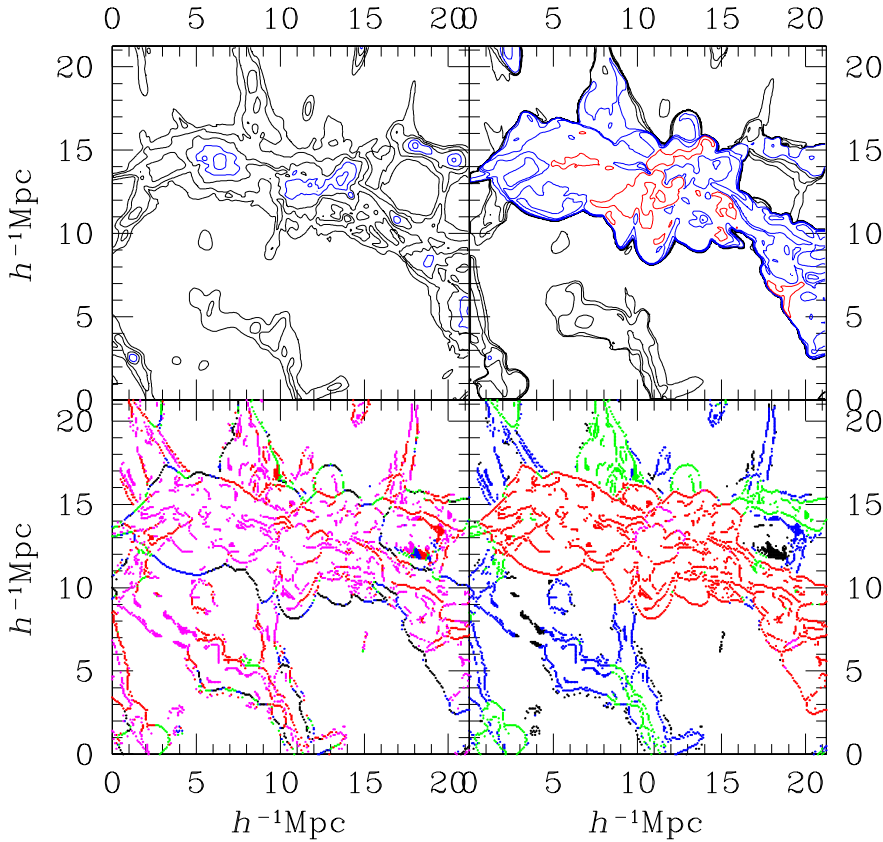}
\caption{Two-dimensional slice around a region containing two
clusters/groups at $z=0$.
Top panels show the distribution of gas density (left) and
temperature (right).
Bottom panels show the locations of the shocks, which are
color-coded according to shock Mach number $M$ (left) or
shock speed $v_s$ (right).
In the left panel, the colors are coded as follows:
black for $M > 100$, blue for $30 < M < 100$, green for $10 < M < 30$,
red for $3 < M < 10$, and magenta for $M < 3$.
In the right panel, the colors are coded as follows:
black for $v_s < 15\ {\rm km\ s^{-1}}$,  blue for
$15 < v_s <65\ {\rm km\ s^{-1}}$, green for $
65 < v_s < 250\ {\rm km\ s^{-1}}$, red for
$250 < v_s < 1000\ {\rm km\ s^{-1}}$, and magenta for
$v_s < 1000\ {\rm km\ s^{-1}}$.}
\end{figure*}

\begin{figure*}
\includegraphics[width=1\textwidth]{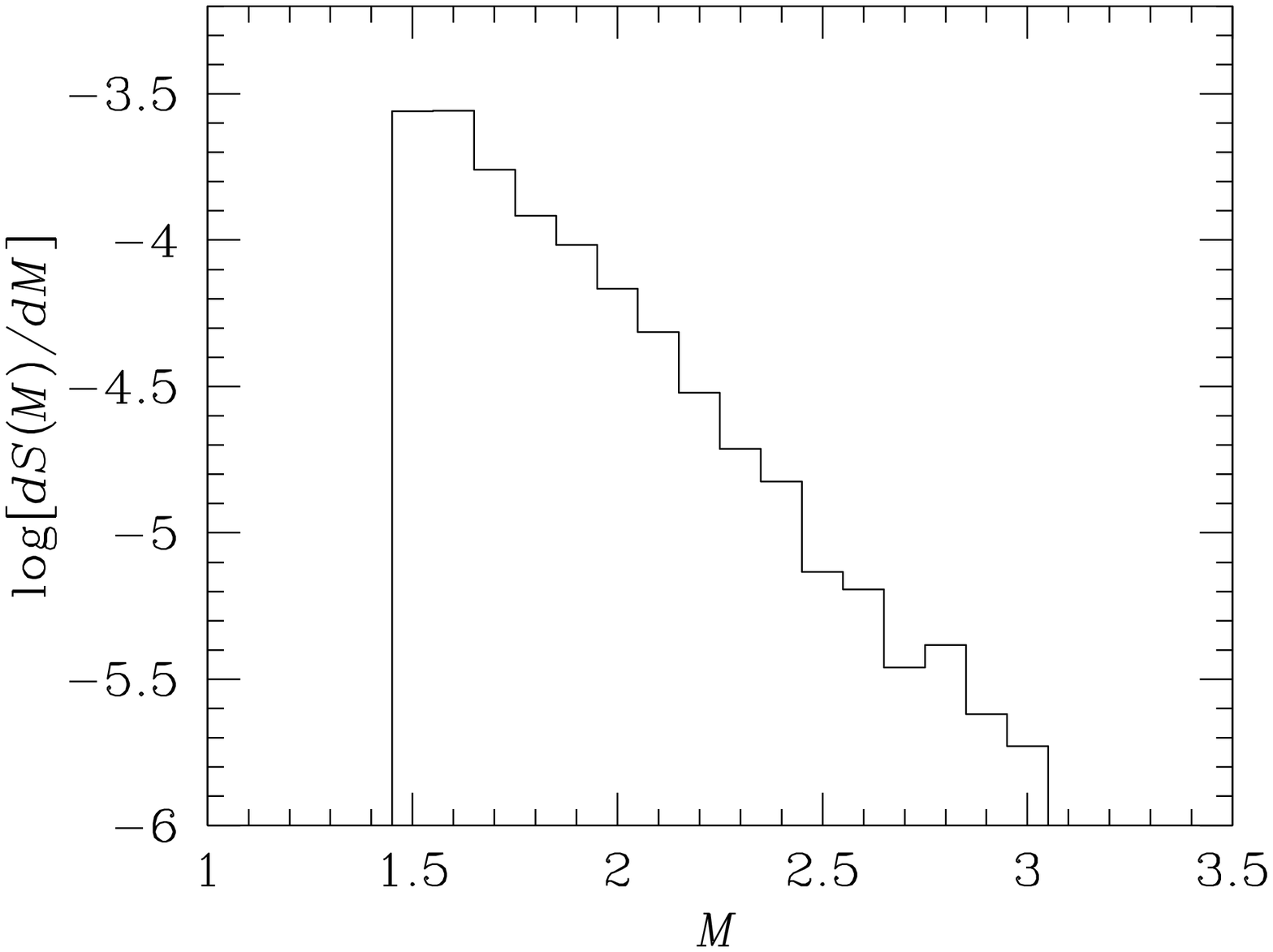}
\caption{Frequency of shocks expected to be found at $z=0$ in cluster
outskirts as a function of $M$. Figure from D. Ryu.}
\label{fig:ryu2}
\end{figure*}

Ryu et al. (2003) performed uniform grid simulations with limited
resolution, in which cluster core regions were not properly resolved,
while surrounding outskirts were fairly well reproduced (see Fig. 6). Hence, the
shocks identified in these simulations are mostly those in cluster
outskirts. A cell is determined to have a shock if it meets the
following requirements: 1. $\nabla\cdot \mathbf{v} < 0$, 2. $ \nabla T
\cdot \nabla s > 0$, 3. $T_2>T_1$ and 4. $\rho_2> \rho_1$.  where
$\mathbf{v}$ is the velocity field, $T$ is the temperature, $\rho$ is
the density, and $s=T/ \rho ^{\gamma -1}$ is, in X-ray astronomy
parlance, the entropy. The subscripts 1 and 2 denote up- and
downstream quantities, respectively.  To find shocks, one loops through
rows of cells along each of the coordinate axes and identifies
one-dimensional shock structures in each direction. Fig.~\ref{fig:ryu2} shows the surface
area, $S$, of identified shocks with the preshock gas temperature $T_1
> 10^7$~K, normalized by the entire volume of the simulation box, at
the present epoch as a function of shock Mach number.  The quantity
$S$ provides a measure of shock frequency. The Mach number of shocks
expected to be found in cluster outskirts is low with $M\ \la\ 3$. The
frequency increases to weakest possible shocks with $M \sim 1$.

\cite{vazza09a} identifies shocks in cosmological simulations using a method based on velocities (instead of temperatures).
They found that the overall differential
distribution of shocks with their Mach number in the cosmic volume is
very steep, with $\beta\sim -1.6$ (with $M dN/dM\propto M^{\beta}$ ),
and the bulk of the detected shocks at any Mach number is found in
low-density regions. The Mach number distribution of detected shocks becomes
increasingly steeper with ambient density: $\beta \sim -
3$ to $- 4$ is found in clusters and their outskirts. They find that
before the epoch of re-ionization, $z > 6$, roughly 30 \% of the
simulated volume is shocked. Then as soon as re-ionization plays a
role, the temperature of the simulated volume increases and the Mach
number distribution of shocks at redshift $z\sim 3 - 6$ undergoes a dramatic
change becoming very steep and dominated by weak shocks.
The bulk of the energy in their simulations is dissipated in galaxy
clusters which contribute about 75 \% of the total
energy dissipation (about 80 \% if the
contribution from cluster outskirts is included), while filaments contribute about
15 \% of the total energy dissipation.

In line with previous numerical studies, relatively weak shocks are
found to dominate the process of energy dissipation in the simulated
cosmic volume, although they find a larger ratio between weak and strong
shocks with respect to previous studies. The bulk of energy is
dissipated at shocks with Mach number $M \sim 2$ and the fraction of strong
shocks decreases with increasing density of the cosmic
environments.


\citet{skillman08} improved the method employed by Ryu et
al. (2003) that can produce errors when examining shocks whose
direction of motion is not oriented along a coordinate axis.  
They find that the Mach number evolution can be
interpreted as a method to visualize large-scale structure
formation. Shocks with $M < 5$ typically trace mergers and complex
flows, while $5 < M < 20$ and $M > 20$ generally follow accretion onto
filaments and galaxy clusters, respectively.

\begin{figure*}
\includegraphics[width=1\textwidth]{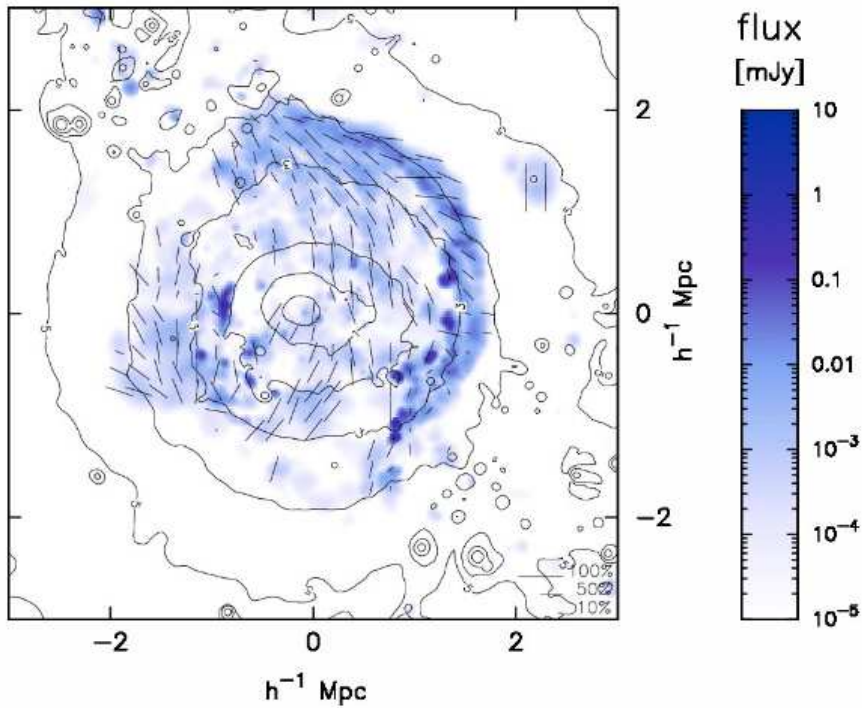}
\caption{Radio emission from a galaxy cluster taken from the Mare
  Nostrum simulation and resimulated at higher resolution. The vectors
  denote the degree and the direction of polarisation of the radio emission. The
  contours indicate the X-ray emission. (From M. Hoeft)}
  \label{fig:hoeft}
\end{figure*}



\subsection{Cosmological simulations with particle acceleration}

In recent years a number of cosmological simulations have started to include some treatment of cosmic ray physics with particle acceleration at shocks.

Using a shock-finding formalism tuned to smoothed-particle hydrodynamics simulations, \citet{hoeft08} have analyzed the
MareNostrum Universe simulation which has $2 \times 1024^3$ particles in a 500
$h^{-1}$ Mpc box. In addition, they have used the formalism
derived in \cite{hoeft07} to produce artificial radio maps of massive
clusters. Several clusters were found to show radio objects with
similar morphologies to observed large-scale radio relics (see Fig.~\ref{fig:hoeft}), whereas about
half of the clusters show only very little radio emission. In
agreement with observational findings, the maximum diffuse radio
emission of galaxy clusters depends strongly on their X-ray
temperature. Moreover, it was found that the so-called accretion
shocks cause only very little radio emission.

Using a particle-mesh and Eulerian hydrodynamics code, Ryu et al. have been studying the gas 
thermalization and CR acceleration efficiencies of shocks in cosmological grid simulations.
In order to quantify the energy dissipation at cosmological
shocks, the incident shock kinetic energy flux, $F_{\rm kin}= (1/2)
\rho_1 v_{s}^3$ where $\rho_1$ is the preshock gas density, and $v_s$ the shock velocity, was
calculated for each shock, and the results are shown in Fig.~\ref{fig:ryu2}.  

\begin{figure*}
\includegraphics[width=1\textwidth]{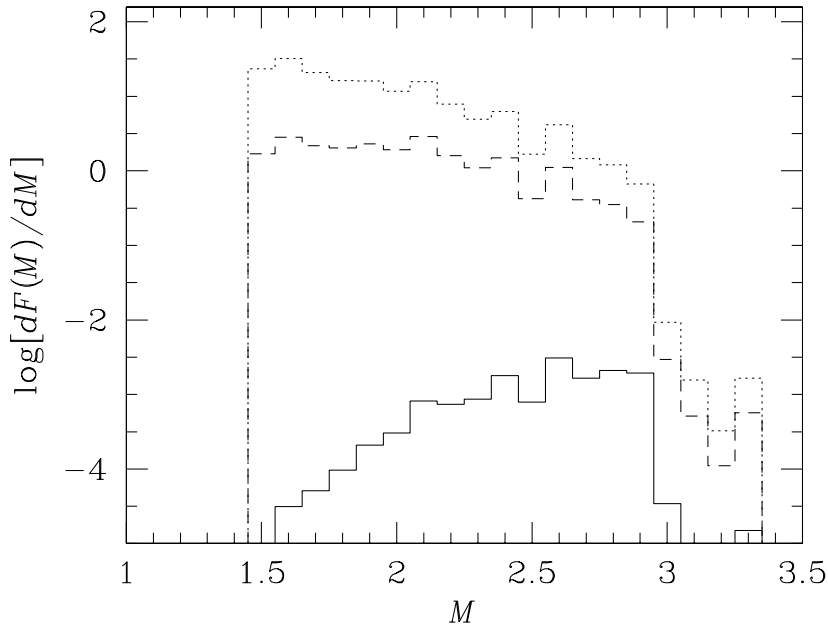}
\caption{Dotted line: Kinetic energy flux per comoving volume passing
through surfaces of shocks in cluster outskirts as a function of $M$
at $z=0$.
Dashed line: Gas thermal energy dissipated at the shocks as a function
of $M$.
Solid line: CR energy accelerated at the shocks as a function of $M$.
All the quantities are in units of
$10^{40}$ ergs s$^{-1}(h^{-1}{\rm Mpc})^{-3}$}
\label{fig:ryu3}
\end{figure*}

Then, the kinetic energy flux through shock surfaces, normalized by the entire volume of
the simulation box, as a function of shock Mach number was
calculated. The resulting kinetic energy flux is shown by the dotted
line in Fig.~\ref{fig:ryu3}. The kinetic energy flux through shock
surfaces is larger for weaker shocks; energetically weaker shocks are
more important processing more shock energy. 
The acceleration efficiency for cosmic rays, $\eta$, depends on the Mach number and the injection
parameter, $\epsilon_B$, which is defined as $\epsilon_B \equiv
B_0/B_{\perp}$, the ratio of the mean magnetic field strength aligned
with the shock normal to the amplitude of the postshock turbulent wave
field. In addition, it is assumed that the CR population is
isotropized with respect to the local Alfv\'enic wave turbulence,
which would in general drift upstream at the Alfv\'en speed with
respect to the bulk plasma.  This reduces the velocity difference
between upstream and downstream scattering centers compared to the
bulk flow, leading to less efficient shock acceleration. Moreover, the dissipation of
Alfv\'en turbulence heats the inflowing plasma in the precursor, which
leads to weakening of the subshock strength. (See \citet{kj07} for
further discussions and references.)

\begin{figure*}
\includegraphics[width=1\textwidth]{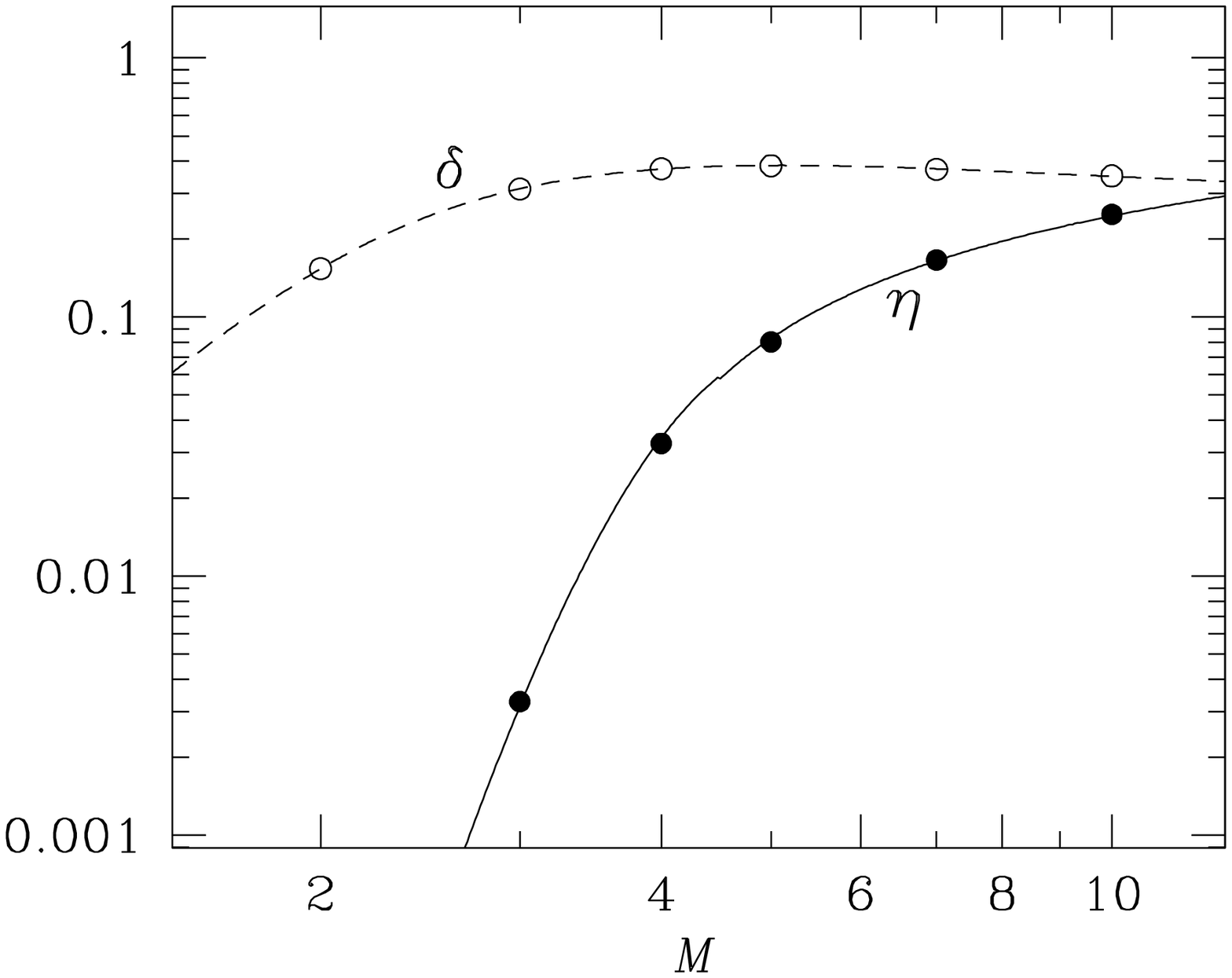}
\caption{Gas thermalization efficiency, $\delta(M)$, and CR
acceleration efficiency, $\eta(M)$, as a function of Mach number.
Open and filled circles are the values estimated from numerical
simulations based on a shock acceleration model and dashed and solid lines are
the fits.}
\label{fig:ryu4}
\end{figure*}

Fig.~\ref{fig:ryu4} shows the gas thermalization and CR acceleration
efficiencies, defined as $\delta(M) \equiv F_{\rm th}/F_{\rm kin}$ and
$\eta(M) \equiv F_{\rm CR}/F_{\rm kin}$, respectively, where $F_{\rm
  th}$ and $F_{\rm CR}$ are the gas thermal and CR energy flux,
respectively, generated at shocks; they are from numerical simulations
of quasi-parallel shocks with speed $v_s=M \cdot 150 {\rm
  km\ s^{-1}}$ ($T_1=10^6$K), and based on a shock acceleration model of \citet{kj07}
with $\epsilon_B = 0.25$.  At strong shocks with $M\ \ga\ 10$, the
injection rate is high enough so that the CR acceleration efficiency
nearly saturates and becomes almost independent of the parameter
$\epsilon_B$. At weak shocks, on the other hand, the level of
injection lies in the regime where the CR acceleration efficiency
increases with the injection rate; consequently, $\eta$ depends
sensitively on $\epsilon_B$. Note that the value of $\eta$ presented
here is $\sim 1/2$ of that presented in \citet{krco07} for
$M\ \la\ 5$, while it remains about the same for stronger shocks.

By adopting the efficiencies in Fig.~\ref{fig:ryu4}, the gas thermal
and CR energy fluxes dissipated at cosmological shocks as a function
of shock Mach number were calculated in the same way the kinetic
energy flux through shock surfaces was calculated. The resulting
thermal and CR energy fluxes are shown with dashed and solid lines in
Fig.~\ref{fig:ryu3}. For the generation of thermal energy, still
weaker shocks are more important; shocks with $M\ \la\ 2.5$ contribute
the most to the thermal energy. On the other hand, the generation of
CR energy is more efficient at shocks with $M\ \ga\ 2$. These indicate
that in cluster outskirts, while shocks with $M\ \la\ 2$ are more
abundant, shocks with $M\ \ga\ 2$ would be more frequently detected as
radio relics. We note that the results in Fig.~\ref{fig:ryu3} are
consistent with those estimated for cosmological shocks in all the
regions of the IGM; while shocks around $M \sim 2$ contribute most to
the generation of gas thermal energy, shocks around $M \sim 3$
contribute most to the generation of CR energy \citep{ryu03,krco07}.

At weak shocks, DSA is known to be rather inefficient and the CR 
pressure remains dynamically insigniÞcant, partly because the injection from thermal to 
nonthermal particles is inefficient (e.g., Kang et al. 2002). Recently, Kang \& Ryu
(2010) found that at weak shocks 
expected to form in clusters much less than $10^{-3}$ 
of particles are injected into CRs and much less than $\sim 1$\% of the shock ram pressure is converted into 
the downstream pressure of CR protons, so the particle acceleration is virtually negligible.

This has been followed up in recent AMR simulations by Vazza (2011 in prep.) that implement 
prescriptions for CR acceleration at shocks. This work suggests that the shock acceleration efficiency needs to be increased in order to explain the occurrence of radio relics. \citet{krco07} have already suggested that that pre-existing populations of cosmic rays can boost the acceleration efficiency significantly, and this may be part of the answer. Thus the observations of cluster radio relics can inform us about the microphysics in weak shocks. This will be the subject of the next section.

\section{Microphysics of shocks}

\subsection{Plasma physics in CR-modified shocks}


The heating processes in collisionless shocks are complex phenomena. The
{\sl irreversible} transformation of a part of the kinetic energy of
the ordered bulk motion of the upstream flow into the energy of the
random motions of plasma particles in the downstream flow in
collisionless shocks is due to the plasma instabilities in the thin
dissipation region where the wave-particle interactions provide
momentum and energy redistribution between different components.

\begin{figure*}
\includegraphics[width=0.8\textwidth]{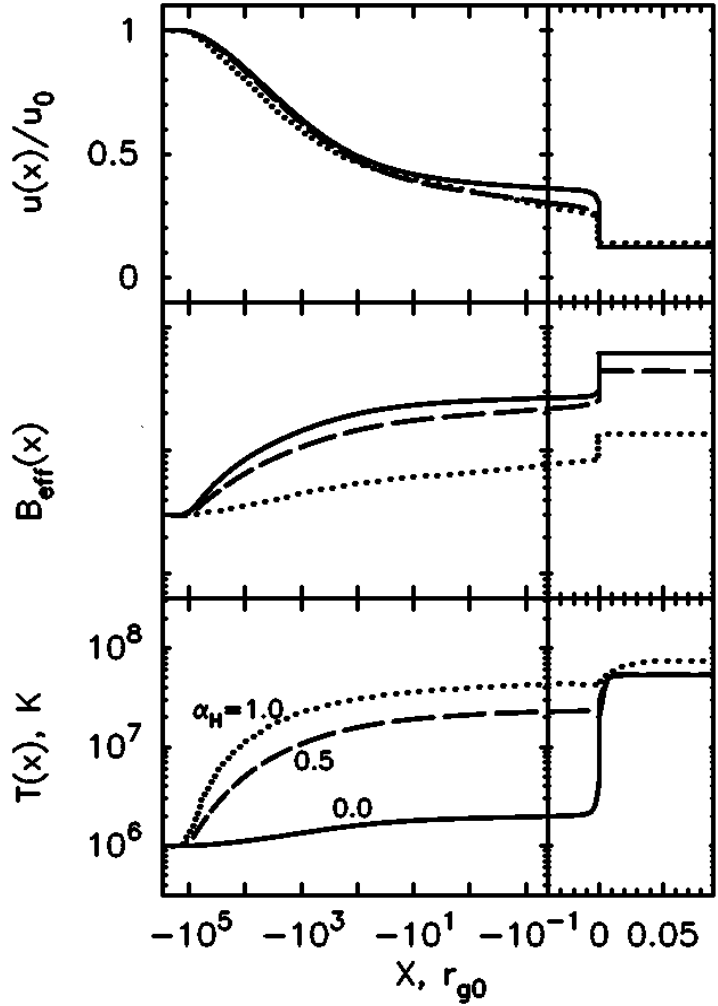}
\caption{Results of non-linear simulations in the case of the far
upstream temperature $T=10^6$~K with different values of turbulence
dissipation parameter $\heatpar$ taken from Vladimirov et al.
(2008). The solid, dashed and dotted lines correspond, respectively,
to $\heatpar = 0$, $0.5$ and $1.0$. The plotted quantities are the
bulk flow speed $u(x)$, the effective amplified magnetic field
$B_{\rm eff}(x)$ and the thermal gas temperature $T(x)$. The shock is
located at $x=0$. There is a change from the logarithmic to the
linear scale at $x=-0.05$. The distances are measured in
relativistic proton gyroradius unit $r_\mathrm{g0}$.}
\label{CRshock}
\end{figure*}

The front of a strong collisionless shock wave may consist of a
precursor and a viscous velocity discontinuity (subshock) of a local
Mach number that is smaller than the total Mach number of the shock
wave (Fig.~\ref{CRshock}). The compression of matter at the subshock
can be much lower than the total compression of the medium in the
shock wave allowing for a high compression in the precursor. We refer
to such shocks as CR-modified.  The standard collisional model of a strong
single-fluid shock predicts a particle temperature $kT
\rightarrow (3/16)\cdot mv_{\rm sh}^2$ for $\gamma_g = 5/3$. However,
this is not generally valid in multi-fluid plasma and CR-modified
shocks since the total shock compression ratio, $R_{\rm t}(v_{\rm sh})$,
depends on the shock velocity, $v_{\rm sh}$, in the strong shock
limit.

In a simplified  steady-state model, a strong CR-modified shock can
be parameterized  by the total Mach number of the shock ${\cal
M}_{\rm tot}$ and the Mach number of the subshock ${\cal M}_{\rm
sub}$. Then the downstream ion temperature $T^{(2)}_{\rm i}$ can be
estimated for the CR modified shock of a given velocity $v_{\rm sh}$
if the total compression ratio, $R_{\rm t}$, is known:
\begin{equation}
T^{(2)}_{\rm i} \approx \phi({\cal M}_{\rm sub})  \cdot \frac{ \mu~
v_{\rm sh}^2}{\gamma_{\rm g}R_{\rm t}^2(v_{\rm sh})} ,~~{\rm
where}~~ \phi({\cal M}_{\rm sub}) = \frac{2 \gamma_{\rm g} {\cal
M}_{\rm sub}^2 - (\gamma_{\rm g} -1)}{(\gamma_{\rm g} -1){\cal
M}_{\rm sub}^2 + 2}. \label{eq:tcr}
\end{equation}
The total compression ratio  $R_{\rm t}$ in Eq.~\ref{eq:tcr} depends
on the precursor heating rate and it is the main parameter in the
steady state CR modified shock to determine the postshock ion
temperature $T^{(2)}_{\rm i}$ \citep[][]{b05,bdd08}.

Shocks may also generate turbulence through cosmic-ray driven instabilities. In the shock
precursor, this turbulence is certain to play a critical role in non-linear models of
strong, CR-modified shocks. Different non-linear approaches to the
modeling of the large-scale structure of a shock undergoing efficient
cosmic ray acceleration
\citep[e.g.,][]{bl01,ab06,veb06,bell04,vbe08,zp08,krj09} have been
proposed. Many of these have predicted the presence of strong MHD
turbulence in the shock precursor.  An exact modeling of the shock structure in a
turbulent medium, including nonthermal particle injection and
acceleration, requires a nonperturbative, self-consistent description
of a multi-component and multi-scale system including the strong
MHD-turbulence dynamics.

A Monte-Carlo model of nonlinear diffusive shock acceleration accounting for magnetic field
amplification through resonant instabilities induced by accelerated
particles, and including the effects of dissipation of turbulence
upstream of a shock and the subsequent precursor plasma heating was
studied by \citet{vbe08}.  Feedback effects between the plasma heating
due to turbulence dissipation and particle injection are strong,
adding to the nonlinear nature of efficient shock acceleration.

Describing the turbulence damping in a parametrized way, \citet{vbe08}
reach two important results: first, for conditions typical of
supernova remnant shocks, even a small amount of dissipated turbulence
energy ($\sim 10\%$) is sufficient to significantly heat the precursor
plasma, and second, the heating upstream of the shock leads to an
increase in the injection of thermal particles at the subshock by a
factor of a few. In their results, the response of the non-linear shock
structure to the boost in particle injection prevented the efficiency
of particle acceleration and magnetic field amplification from
increasing. The non-linear models models with turbulence generation
and dissipation may lead to a scenario in which particle injection
boost due to turbulence dissipation results in more efficient
acceleration and even stronger amplified magnetic fields than without
dissipation. In Fig.~\ref{CRshock} the effects are illustrated for a
stationary, non-linear Monte-Carlo simulation of a forward shock in a
supernova remnant of velocity of 5,000~$\kms$ accelerating particles
up to maximal energies of about 10$^{5}$~GeV \citep{vbe08}. The shock
velocity is higher than is expected for cluster accretion shocks but
the simulation can be used to investigate the qualitative behavior and
to obtain the appropriate scalings. The maximal energy depends on the
ratio of the shock scale size and particle escape physics that is
simply modeled by a free escape boundary in the Monte Carlo
simulation.  Ultra-high energy CR acceleration above 10$^9$~GeV by
cluster accretion shocks was shown by \citet{nma95} and \citet{krj96}
to be a plausible scenario. In that case the scale size of CR
precursor with magnetic field amplification can be about 100~kpc.

\citet{vbe08} demonstrated that the effect of turbulent
dissipation on the thermal plasma is evident in the values of the
pre-subshock temperature $T_1$, the downstream temperature $T_2$,
and the volume-averaged precursor temperature $\left< T(x<0)
\right>$ (the averaging takes place over the CR precursor region).
The temperatures were calculated from the thermal particle pressure
using the ideal gas law. The value of the pre-subshock
temperature $T_1$ depends drastically on the level of the turbulent
dissipation $\heatpar$, increasing from $\heatpar=0$ to
$\heatpar=0.5$ by a factor of 11 in the case of $T=10^6$~K. It is
less in the case of  $T=10^6$~K than for  $T\sim10^4$~K because
the efficiency of the CR streaming instability in generating the
magnetic turbulence is less for the smaller Mach number.
The values of the temperature as high as $T_1$ are achieved upstream
only near the subshock; the volume-averaged upstream temperature,
$\left< T(x<0) \right>$, is significantly lower. The factor by which
the average temperature $\left< T(x<0) \right>$ increases in case of
$T\sim10^6$~K is about 2.3.
The multi-fluid processes described above can preheat the
accreting gas. An important prediction of the models of
strong shocks is the possibility to amplify an initial seed magnetic
field by a few orders of magnitude.  CR currents and CR pressure
gradients upstream of the strong shock can drive magnetic fluctuations
on the shock precursor scale length. The
precursor scale size is $10^9$ times larger than the subshock
transition region where strong, small-scale magnetic field
fluctuations are directly produced by instabilities of super-Alfvenic
bulk plasma flows. These small-scale fluctuations are responsible for
bulk dissipation and the adiabatic amplification of the transverse
magnetic field in collisionless shocks.  From Fig.~\ref{CRshock}, it
is clear that the magnetic field amplification factor depends on the
non-linear processes of particle injection through turbulent
dissipation in the shock precursor.

Above, we discussed the ion temperature in multi-fluid shocks.
However, cluster X-ray spectra depend primarily on the electron
temperature, while MHD-type wave dissipation in the shock precursor
may preferentially heat ions.  However, the complete $e-i$ Coulomb
equilibration requires the system age $\tau_{\rm ei} \gsim
10^{10}~T_6^{3/2}/n$, where the postshock density $n$ is measured in
$\cmc$, the ion temperature $T_6$ is in 10$^6$ K, and $\tau_{\rm ei}$
is in seconds \citet{mewe90}. Since the number density around the
cluster virial radius is typically above 10$^{-5},\cmc$, the
temperature equilibration scale length in the accretion shock
precursor is below 100 kpc. The electron temperature downstream of the shock depends on collisionless electron heating
e.g. \citep{bu99,rlg08,bpp08}.  Non-resonant interactions of the
electrons with strong nonlinear fluctuations generated by kinetic
instabilities of the ions in the transition region inside the shock
front may play the main role in the heating and pre-acceleration of
the electrons, as it was shown in the model by \citet{bu99}. They
calculated the electron energy spectrum in the vicinity of shock waves
and showed that the heating and pre-acceleration of the electrons
occur on a scale of the order of several hundred ion inertial lengths
$l_i$ in the vicinity of a viscous discontinuity. Although the
electron distribution function is significantly out of equilibrium near
the shock front, its low energy part can be approximated by a
Maxwellian distribution. 

The effect of the non-adiabatic electron heating efficiency,
$\beta_{e}$, on the degree of non-equipartition was
studied recently by \citet{ws09}. They
have shown that for a cluster with a mass of $M_{\rm vir} \sim 1.2
\times 10^{15} \Msun$, electron and ion temperatures differ by less
than a percent within the virial radius. The difference is 20\% for
a non-adiabatic electron heating efficiency of  $1/1800$ to 0.5 at
$\sim 1.4$ of the virial radius. Beyond this radius, the
non-equipartition effect depends rather strongly on $\beta_{e}$ \citep{ws09}, and such a strong dependence at the shock radius can be used to distinguish shock heating models or constrain the shock heating efficiency of electrons. 

As opposed to semi-analytic kinetic theory methods \citep{achterberg01, keshet05} or Monte Carlo test-particle simulations \citep{niemiec04, ellison04}, 
PIC simulations can tackle the problem of particle acceleration in shocks without the need for simplifying assumptions about the nature of the magnetic turbulence 
or the details of wave-particle interactions. 
Multi-dimensional PIC simulations of relativistic un-magnetized shocks have been presented, e.g. by \cite{spitkovsky08} for electron-ion flows. \citet{kt10} present a two-dimensional
particle-in-cell simulation to investigate weakly magnetized
perpendicular shocks.  The simulated thermal energies of electrons and
ions in the downstream region are not in equipartition but their
temperature ratio $\beta_{e} = T_e / T_i \sim$ 0.3 - 0.4 is high
enough indicating rather efficient non-adiabatic heating of the electrons
in the shock transition region.

Using 2.5D PIC simulations, \cite{sironi10} find that in ÒsubluminalÓ shocks, where relativistic particles can escape 
ahead of the shock along the magnetic field lines, ions are efficiently accelerated via a Fermi-like 
mechanism. The 
scattering is provided by short-wavelength non-resonant modes produced by Bell's instability (Bell 
2004), whose growth is seeded by the current of shock-accelerated ions that propagate ahead of the 
shock. Upstream electrons enter the shock with lower energy than ions, so they are more strongly tied to the field. As a result, only about 1\% of the incoming 
electrons are Fermi-accelerated at the shock before they are advected downstream, where they populate 
a steep power-law tail. Thus, efficient electron heating is the universal property of relativistic electron-ion shocks, but signiÞcant nonthermal acceleration of electrons is hard to achieve in magnetized flows and requires weakly magnetized shocks, where magnetic fields self-generated via the Weibel instability are stronger than the background field.


\subsection{Entropy in CR-modified Accretion Shocks}

A distinctive feature of CR-modified shocks is their high gas
compression factor $R_{\rm t}(v_{\rm sh})$ that can be much higher
than the single fluid shock limit of $(\gamma_{\rm g} +1)/(\gamma_{\rm
  g}-1)$. Since the postshock gas entropy for a
strong multi-fluid shock scales as $R_{\rm t}(v_{\rm
  sh})^{-(\gamma_{\rm g} +1)}$, it can be significantly reduced
compared to the single-fluid adiabatic shock of the same velocity
\citep[\emph{e.g.,} ][]{bdd08}. The effects are due to energetic
particle acceleration and magnetic field generation. Energetic
particles can penetrate into the shock upstream region. They are
coupled with the upstream gas through fluctuating magnetic fields
(including the \alf waves). Magnetic field dissipation provides gas
preheating and entropy production in the shock precursor. Energetic
particles deplete the momentum distribution in the gas resulting in a higher gas
compression and a reduced temperature.  In Fig.~12 the
post-shock gas entropy is shown as a function of the fraction
$\eta_{\rm esc}$. The curves were calculated for a strong shock with
no precursor heating i.e.  $\heatpar=0$.

\begin{figure*}\label{entropy}
\includegraphics[width=1\textwidth]{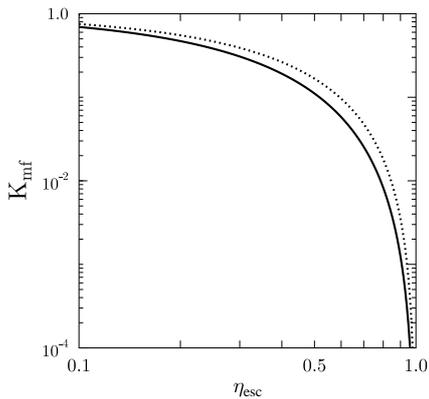}
\caption{Normalized post-shock gas entropy, $K_{\rm mf} = K_0\,
T/\rho^{(\gamma_{\rm g} -1)}$, as a function of the
CR particle energy escape flux $Q_{\rm esc}$ carried by energetic
particles. The dimensionless flux is defined as $\eta_{\rm esc}= Q_{\rm esc}/\Pi_{\rm kin}$,
 where $\Pi_{\rm kin} = \rho_1 v_{\rm sh}^3/2$. The upper
curve (dotted) corresponds to an effective adiabatic exponent
$\gamma_{\rm g}$ = 4/3 (relativistic gas), while the lower (solid) curve
corresponds to $\gamma_{\rm g}$ = 5/3. The postshock entropy
is normalized to $K_{\rm mf}(\eta_{\rm esc}=0)$.}
\end{figure*}

In galaxy clusters, entropy profiles have been measured by \citet{prattea10} at least out to $R_{1000}$ in 31 nearby galaxy clusters from the \emph{XMM-Newton} cluster structure survey (REXCESS) and out to $R_{500}$ in thirteen systems. The effects of CRs on the entropy production in shocks illustrated above can be of importance for  accretion shocks that are expected to be located at larger radii. It will require deeper exposures of the next generation telescopes to measure the entropy profiles at these radii. The entropy in the inner cluster regions can be also affected by internal shocks if the shocks are efficient accelerators of CRs. The current entropy profile simulations \citep[e.g.][]{borganiea08,bk09} ought to be extended to include CR physics.           

\section{Summary}

It is only now, with low-frequency radio telescopes, long exposures with high-resolution X-ray satellites and $\gamma$-ray telescopes, that we are beginning to learn about the physics of cluster outskirts. In the coming years SZ telescopes are going to deliver further insights into the plasma physics of these special regions in the Universe. The last years have already shown tremendous progress with detections of shocks, estimates of magnetic field strengths and constraints on the particle acceleration efficiency. The main points of this review are listed below:

\begin{itemize}

\item Cosmological shocks are collisionless: Coulomb collisions are not
sufficient to provide the viscous dissipation of the incoming flow,
and collective effects play a major role.

\item Collisionless shocks generate energetic, non-thermal particles that
can penetrate far into the upstream flow. The particles decelerate the
flow and preheat the gas. They can also efficiently generate strong
fluctuating magnetic fields in the upstream region. This turbulence,
which may result from cosmic-ray driven instabilities, in the shock
precursor, is certain to play a critical role in non-linear models of
strong, CR-modified shocks.

\item A few merger shocks have been identified in X-rays, exhibiting both a sharp gas density edge and an
unambiguous temperature jump: in the "bullet cluster", 1E 0657Ð56 \citep{markevitch02}, A520 \citep{markevitch05}
and Abell 2146 \citep{russell10}.

\item On the basis of X-ray observations, microphysical properties of shocks can be derived. Markevitch et al. (2006) find that the temperatures across merger shocks are consistent with instant
heating; equilibration between electrons and protons on the collisional timescale is excluded. 
The electron temperature downstream of the shock depends on collisionless electron heating
e.g. \citep{bu99,rlg08,bpp08}.  Non-resonant interactions of the
electrons with strong nonlinear fluctuations generated by kinetic
instabilities of the ions in the transition region inside the shock
front may play the main role in the heating and pre-acceleration of
the electrons. Future X-ray observations extending to the virial radius or even close to the shock radius should be able to detect signatures of non-equipartition at shocks.

\item \emph{Suzaku} observations provide evidence for departures from hydrostatic equilibrium around
and even before the virial radius as it is seen in the galaxy cluster
Abell 1795 \citep[\emph{e.g.,} ][]{bautzea09}.  Evidence for nonthermal pressure
support suggests that bulk motions from mergers could be
making a significant contribution to the gas energy in the outskirts
of this cluster \citep{georgeea09}. 

\item Recently discovered radio relics provide very strong evidence for shock acceleration at merger shock waves in galaxy clusters.
Van Weeren et al. (2010) find that on average smaller radio
relics have steeper spectra. Such a trend is in line with predictions from
shock statistics derived from cosmological simulations
\citep{skillman08, battaglia09,hoeft08}. They find that larger shock
waves occur mainly in lower-density regions and have larger Mach
numbers, and consequently shallower spectra. 

\item Observations of radio relics in combination with X-ray observations suggest magnetic field strengths in relics of $>3$ \muG\, and efficiencies of 0.2\% of the kinetic energy dissipated in a shock going into relativistic electrons (Finoguenov et al. 2009, van Weeren et al. 2010) .

\item Radio relics also suggest that the efficiency of shock acceleration at weak shocks is higher than predicted in theories of diffusive shock acceleration. 

\item The bulk of the energy in their simulations is dissipated in galaxy
clusters which contribute about 75 \% of the total
energy dissipation (about 80 \% the
contribution from cluster outskirts is included), while filaments contribute about
15 \% of the total energy dissipation. In
agreement with observational findings, the maximum diffuse radio
emission of galaxy clusters depends strongly on their X-ray
temperature. While shocks around $M \sim 2$ contribute most to
the generation of gas thermal energy, shocks around $M \sim 3$
contribute most to the generation of CR energy \citep{ryu03,krco07}.

\item The effects of CRs on entropy production in shocks as illustrated above can be of importance for accretion shocks that are expected to be located at larger radii. It will require deeper exposures of the next generation telescopes to measure the entropy profiles at these radii.

\end{itemize}
Many open questions remain. So far still little is known about shock acceleration
and magnetic field generation in clusters, nor about the microstructure of shocks in very dilute plasmas.

Most likely, all these processes are related and their understanding requires input from, both, observations and theory. The next years will see a rapid growth in observational data. On the theoretical side, we expect substantial progress from new techniques that simulate particle acceleration in a magnetohydrodynamical setting. Fully kinetic particle-in-cell simulations provide a powerful tool for exploration of the structure of collisionless shocks from first principles, thus determining 
self-consistently the interplay between shock-generated waves and accelerated particles \citep{sironi10}. Results from these simulations can then serve as input for cosmological simulations that include CR-physics.

\begin{acknowledgements}
We thank an anonymous referee for a very careful read of our manuscript.
MB acknowledges support by the DFG Research Unit "Magnetisation of
Interstellar and Intergalactic Media: The Prospects of Low-Frequency
Radio Observations". DR was supported in part by R\&D Program through
the National Fusion Research Institute of Korea funded by the
Government funds. A.M.B. was supported in part by RBRF grants 09-02-12080, 11-02-00429a, by the RAS
Presidium Programm, and the Russian government grant 11.G34.31.0001
to Sankt-Petersburg State Politechnical University. He performed some of
the simulations at the Joint Supercomputing Centre (JSCC RAS) and the
Supercomputing Centre at Ioffe Institute, St.Petersburg. We thank Aurora Simionescu, 
Matthias Hoeft and Franco Vazza for helpful comments on the manuscript.  Finally, the
authors would like the convenors of the workshop in Bern for their
initiative and excellent work and the staff at ISSI for their support.

\end{acknowledgements}

\bibliographystyle{aps-nameyear}      
\bibliography{cluster,marcus}

\end{document}